\documentclass[cite,aps,prl,amsmath,amsfonts,superscriptaddress,amssymb,twocolumn]{revtex4}

\usepackage{graphicx}

\newcommand{\mean}[1]{\overline{#1}}

\newcommand{\citeasnoun}[1]{Ref.~\onlinecite{#1}}

\begin{document}

\title{Efficient generation of correlated random numbers \\ using
Chebyshev-optimal magnitude-only IIR filters}

\author{Alejandro Rodriguez}
\email{alexrod7@mit.edu}
\affiliation{Department of Physics, Massachusetts Institute of
Technology, Cambridge, MA 02139} \author{Steven G. Johnson}
\affiliation{Department of Mathematics, Massachusetts Institute of
Technology, Cambridge, MA 02139}

\begin{abstract}
  We compare several methods for the efficient generation of
  correlated random sequences (colored noise) by filtering white noise
  to achieve a desired correlation spectrum.  We argue that a class of
  IIR filter-design techniques developed in the 1970s, which obtain
  the global Chebyshev-optimum minimum-phase filter with a desired
  magnitude and arbitrary phase, are uniquely suited for this problem
  but have seldom been used.  The short filters that result from such
  techniques are crucial for applications of colored noise in physical
  simulations involving random processes, for which many long random
  sequences must be generated and computational time and memory are at
  a premium.
\end{abstract}

\maketitle

The generation of correlated random sequences, or ``colored noise'',
is important for many physical simulations involving random
processes~\cite{Fox88,Billah90,Kasdin95,Luo04:thermal,Traulsen04}, and
often the required computational time and memory is a critical
concern.  Although many filter-based techniques have been applied to
this
problem~\cite{Fox88,Billah90,Kasdin95,Chan81,Manella89,Garcia92,Young00,Lu05,Komninakis04},
in this brief manuscript we point out that these methods are
suboptimal: a global Chebyshev-optimum stable IIR filter for this
problem may be designed based on techniques developed in the
1970s~\cite{Dudgeon74,Rabiner74}.

Colored-noise generation is required for many types of numerical
simulations, such as
thermodynamics~\cite{Billah90,Kasdin95,Luo04:thermal}, laser noise and
first-passage time problems~\cite{Fox88}, and chaotic
dynamics~\cite{Traulsen04}. In general, any numerical model involving
stochastic differential equations in which there is some background
distribution, nonlinearity, or external noise associated with the
quantities driving the fluctuations will require the use of colored
noise~\cite{Fox88,Kasdin95,Traulsen04,Garcia92,Lu05}.  Such colored
noise is most commonly described by a \emph{linear process}: the
output of a time-invariant linear filter applied to white
noise~\cite{Brockwell02}.  The key implementation question, as we
discuss below, is to determine how to apply this filter process.  With
rare exceptions~\cite{Wio03}, the underlying random process is almost
always Gaussian in nature, if it arises by a physical process governed
by the central-limit theorem~\cite{Reif:stat}.  It is this nearly
ubiquitous case that we focus on here, in which only the correlation
function of zero-mean colored noise is of concern, and not any
higher-order statistics.

From a computational standpoint, the central problem is that many of
these applications require many random sequences to be generated in
parallel, and/or very long sequences, imposing performance and memory
constraints on the noise generation.  This occurs, for example, in the
generation of power-law ($1/f^\alpha$) ``pink noise'' for applications
ranging from hydrology to music, where very long sequences (whose
length may not be known \textit{a priori}) are often required
\cite{Kasdin95}.  Multiple long random sequences also occur in the
generation of Rayleigh random variates for simulation of nonisotropic
scattering, macrocellular systems, and physical models of mobile
multipath radio channels~\cite{Young00,Komninakis04,Baddour05}.
Perhaps the most stringent requirements for colored-noise generation
occur for spatio-temporal noise: where a different random sequence may
need to be generated simultaneously for every point in space.  This
occurs, for example, in modeling of Brownian motion~\cite{Traulsen04},
or for simulation of thermal
radiation~\cite{Luo04:thermal}.\endnote{In the specific problem
considered in \citeasnoun{Luo04:thermal}, the correlation function
could be renormalized to an uncorrelated sequence due to the linearity
of the system, but this trick would not work in more general nonlinear
or nonequilibrium systems.}

\begin{figure}
\begin{center}
\includegraphics[width=0.5\textwidth]{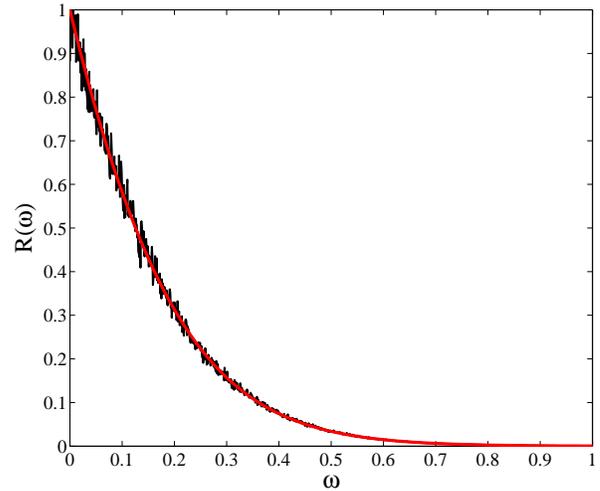}
\end{center}
\caption{Plot of Planck distribution $R(\omega) = a\omega /
(\exp(a\omega)-1)$ vs. $\omega$ (red). The periodogram (spectrum) of a
finite-length correlated random sequence generated by the methods in
this paper is also shown (black), computed by Bartlett's
method~\cite{Oppenheim99}. }
\label{fig:pfunc}
\end{figure}

The fundamental technique for generating correlated random sequences
is to start with white noise (uncorrelated random numbers) and to
apply a filter whose power spectrum matches the desired correlation
spectrum~\cite{Chan81,Billah90,Kasdin95,Komninakis04,Lu05}.  That is,
suppose that we want colored-noise $y_n$ with some desired correlation
function $R_m = \mean{y_n y_{n+m}}$, the (discrete-time)
Fourier transform of which is the correlation spectrum $R(\omega)$.
We then start with white noise $x_n$, usually uncorrelated Gaussian
random numbers, with zero mean and $\mean{x_n^2} = 1$, and
apply a filter $H(\omega)$: in frequency domain, $Y(\omega) =
H(\omega) \, X(\omega)$.  The desired correlation spectrum is achieved
if $|H(\omega)|^2 = R(\omega)$.

Although this filtering operation can be performed entirely in the
frequency domain via a fast Fourier transform (FFT) of the data
sequences~\cite{Billah90,Kasdin95,Lu05}, such an approach is often too
inefficient. Filtering entirely in the frequency domain requires the entire
data sequence $y_n$ to be computed and stored in advance, and if many
long sequences are required the storage becomes prohibitive.  The
alternative is to perform the filtering in the time domain, using
either finite-impulse response (FIR) filters or infinite-impulse
response (IIR) filters (also called recursive or ARMA filters).  Since the
generation of colored noise is of interest to many researchers without
a background in signal processing, let us briefly review the basics of
FIR and IIR filtering, which are covered in more detail by numerous
textbooks, e.g. ~\citeasnoun{Oppenheim99}.  FIR filters consist of a
finite-length sequence $b_n$ that is convolved with the $x_n$.  The
Fourier transform of the $b_n$ is the filter $H(\omega)$, but in
general a finite-length sequence can only approximate an arbitrary
desired spectrum $R(\omega)$.  In particular, an FIR filter yields a
spectrum $H(\omega)$ which is a polynomial in $z=e^{i\omega}$ with
coefficients $b_n$.  A better approximation may be obtained by
generalizing to a ratio of two polynomials (i.e., rational functions),
which leads to IIR filters.  An IIR filter is determined by two
finite-length sequences, $a_n$ ($n=0 \ldots N$, $a_0=1$) and $b_n$ ($n
= 0 \ldots M$), that determine $y_n$ via the recurrence (which can be
written in several equivalent forms):
\begin{equation}
y_n = \sum^{M}_{k=0} b_k x_{n-k} - \sum^{N}_{k=1} a_k y_{n-k} \, .
\label{eq:iir-recurrence}
\end{equation}
That is, $y_n$ is a convolution of $b_n$ with $x_n$ and of $a_n$ with
the previous values of $y_n$.  The filter design problem is then,
given filter orders $N$ and $M$, to find the $a_n$ and $b_n$ that best
approximate the desired spectrum.

Therefore, we must choose what type of filter to apply (IIR or FIR)
and a filter-design method.  Several choices have been previously
proposed in the context of colored-noise generation.  The simplest
method, as we mentioned above, is to just perform a fast Fourier
transform (FFT) of the entire desired spectrum multiplied by random
phases~\cite{Billah90,Kasdin95,Young00,Lu05}.  This is equivalent to
an FIR filter of the same length as the data, designed by the
``window'' method~\cite{Oppenheim99}.  One can also employ FIR filters
of shorter lengths, designed by a variety of standard methods such as
Parks-McClellan~\cite{Rabiner75,Oppenheim99}.  For certain noise
problems, an FIR filter may also be designed by analytical
methods~\cite{Kasdin95,Young00}.  In order to shorten the length of
the required filter, and thus the memory and time requirements, IIR
recursive filters have been
proposed~\cite{Kasdin95,Chan81,Komninakis04}.  In general, the design
of IIR filters is a difficult problem~\cite{Parks87,Oppenheim99}, and
past approaches to colored noise generation by IIR filtering have used
local-optimization~\cite{Komninakis04} or Yule-Walker
methods~\cite{Chan81} that are not guaranteed to yield the
global-optimum filter coefficients.  Another difficulty is that not
all IIR-filter design techniques are guaranteed to yield a stable
filter (one which does not lead to a diverging process). One important
exception is exponentially correlated noise, for which an exact
first-order IIR filter is known analytically and is commonly used
(although typically derived from a stochastic differential equation
$y' = -a y + x$ and not recognized as an IIR filter \textit{per
se})~\cite{Fox88,Manella89,Garcia92,Traulsen04}.  However, there is
a key property of the IIR filter design problem for colored-noise
generation that makes optimal filter design practical: the phase of
the filter $H(\omega)$ is irrelevant, since it is multiplied in any
case by white noise $X(\omega)$ with a random phase.

In particular, we can exploit results by Dudgeon~\cite{Dudgeon74} and
Rabiner~\cite{Rabiner74}, who demonstrated that the \emph{global}
Chebyshev-optimum magnitude-only minimum-phase stable IIR filter-design
problem can be efficiently solved by a sequence of linear-programming
problems~\cite{Crosara83}.  Alternative methods with similar
properties have also been proposed~\cite{Alkhairy95,Zhang96}, and the
Dudgeon and Rabiner technique was recently generalized to
multidimensional IIR filters~\cite{Gorinevsky06}. However, its
applicability to the problem of colored-noise generation does not seem
to have been appreciated, and in this manuscript we demonstrate that it
can yield dramatically more efficient filters than previous approaches.

To demonstrate the efficacy of the various filter-design approaches
for colored-noise generation, we consider an example drawn from
thermodynamic simulations of gray-body thermal
radiation~\cite{Luo04:thermal}. In this problem, thermal effects are
modeled as random fluctuating current sources everywhere in space,
with a correlation spectrum~$R(\omega)=a\omega / (\exp(a\omega)-1)$,
for a constant $a$ determined by the temperature, based on the Planck
distribution.  This distribution is shown in Fig.~1, along with the
periodogram of a finite-length correlated random sequence generated by
the methods in this paper.

An IIR filter is defined by a rational polynomial function in $z=e^{i\omega}$:
\begin{equation}
H(\omega) = \frac{\displaystyle \sum^{M}_{k=0} b_k e^{i\omega
k}}{1 + \displaystyle\sum^{N}_{k=1} a_k e^{i\omega k}}
\label{eq:iir-H}
\end{equation}
where $N$ and $M$ define the filter order and $a_0 = 1$ for
convenience.  A stable minimum-phase IIR filter has all of its poles
and zeros within the unit circle in $z$ (i.e., for $\mathrm{Im}\, \omega >
0$)~\cite{Oppenheim99}. Given a sequence of uncorrelated random
numbers $x_n$, the output colored-noise sequence $y_n$ is then given
by the recurrence (Eq. 1) above.  The required memory, along with the
computation time per output, is therefore $O(N + M)$.  An FIR filter
is the special case $N=0$.  A sequence $y_n$ of length $K$ from an FIR
filter can be generated in $O(K \log M)$ time instead of $O(K M)$ by
use of fast Fourier transforms (via overlap-add or overlap-save
techniques~\cite{Oppenheim99}), but the memory requirements are not
improved.

\begin{figure}
\begin{center}
\includegraphics[width=0.5\textwidth]{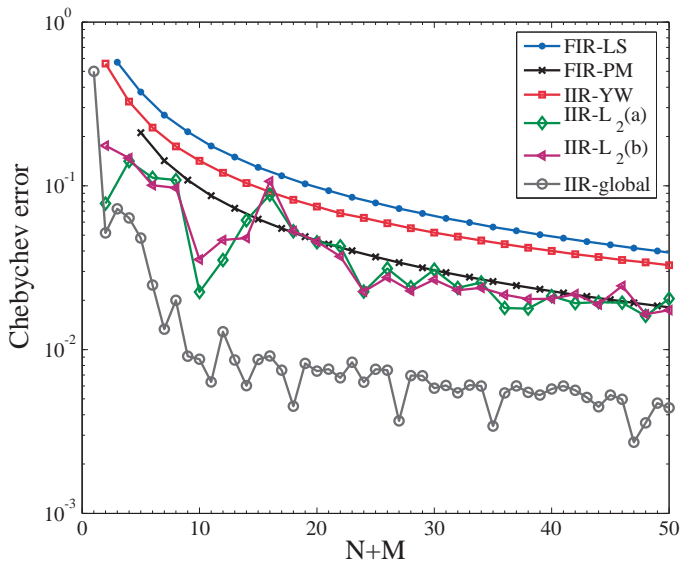}
\caption{Plot of Chebychev error, $L_{\infty} = \max_\omega |R(\omega)
  - |H(\omega)|^2|$, vs. $N+M$ for the Planck distribution $R(\omega)$
  given in Fig.~1. The Chebychev error is plotted for six
  different filter methods: IIR Yule-Walker (red squares), FIR
  Parks-McClellan (black crosses), FIR Least-Squares (blue dots), IIR
  global-optimum filter (gray circles), and two nonlinear
  conjugate-gradient methods optimizing two different $L_2$ norms:
  (a) $L_2(|H|-\sqrt{R})$ (green diamonds) and
  (b) $L_2(|H|^2-R)$ (magenta triangles).
}
\end{center}
\label{fig:abserr}
\end{figure}

In Fig.~2, we plot the $L_\infty$ (Chebyshev) error, $\max_\omega
\left| R(\omega) - |H(\omega)|^2 \right|$, as a function of the total
filter order $N+M$, for filters $H(\omega)$ designed by several
techniques.  (Here, we employ the Chebyshev error over the entire
frequency bandwidth; for other applications, only a subset of the
bandwidth may be of interest.)  The best method, i.e. smallest
\emph{Chebychev error} for any given-order, is the optimal
magnitude-only IIR filter design (implemented using the
differential-correction algorithm~\cite{Dudgeon74,Crosara83}). The
other design methods plotted consist of two FIR and two IIR filter
techniques.  The two linear-phase FIR filter techniques are (from the
Matlab signal-processing toolbox): first, the Parks-McClellan
algorithm, which finds the global Chebyshev
optimum~\cite{Oppenheim99,Rabiner75} (black); second, a least-squares
FIR optimization (blue), as described in \citeasnoun{Parks87}.  The
two IIR filter designs plotted are: first, a nonlinear
conjugate-gradient minimization of a least-squares norm proposed
by~\citeasnoun{Steiglitz70}, and suggested
by~\citeasnoun{Komninakis04} for use in generation of
Rayleigh-correlated noise (green diamonds); second, another
(non-global) optimization technique based on the modified Yule-Walker
algorithm in the Matlab signal-processing
toolbox~\cite{Chan81,Friedlander84}.  The conjugate-gradient method
only finds a local optimum, and is highly sensitive to the starting
point of the optimization for this problem~\cite{Komninakis04}.  Here,
we use the Yule-Walker IIR filter as the starting point, and conjugate
gradient is able to improve upon it significantly (in fact, it happens
to nearly match the Parks-McClellan FIR performance).  We examined the
conjugate-gradient minimization of two different $L_2$ norms
[$L_2(x_{n=1 \ldots N}) = \sqrt{\sum_n x_n^2 / N}$] of the error:
$L_2(|H| - \sqrt{R})$~\cite{Komninakis04,Steiglitz70} and
$L_2(|H|^2-R)$, although it turns out to make little difference in the
result.

\begin{figure}
\begin{center}
\includegraphics[width=0.5\textwidth]{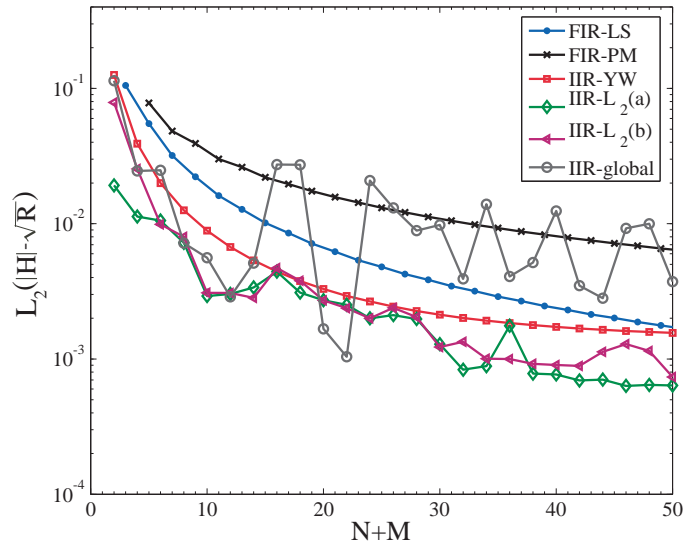}
\caption{Plot of $L_2$ error, $L_2(|H|-\sqrt{R})$,
    vs. $N+M$ for the Planck distribution $R(\omega)$ given in
    Fig.~1. The $L_2(|H|-\sqrt{R})$ error is plotted for
    six different filter methods: IIR Yule-Walker (red squares), FIR
    Parks-McClellan (black crosses), FIR Least-Squares (blue dots),
    IIR global-optimum filter (gray circles), and two nonlinear
    conjugate-gradient methods optimizing two different $L_2$ norms:
    (a) $L_2(|H|-\sqrt{R})$ (green diamonds) and
    (b) $L_2(|H|^2-R)$ (magenta triangles).}
\end{center}
\label{fig:l2a}
\end{figure}

\begin{figure}
\begin{center}
\includegraphics[width=0.5\textwidth]{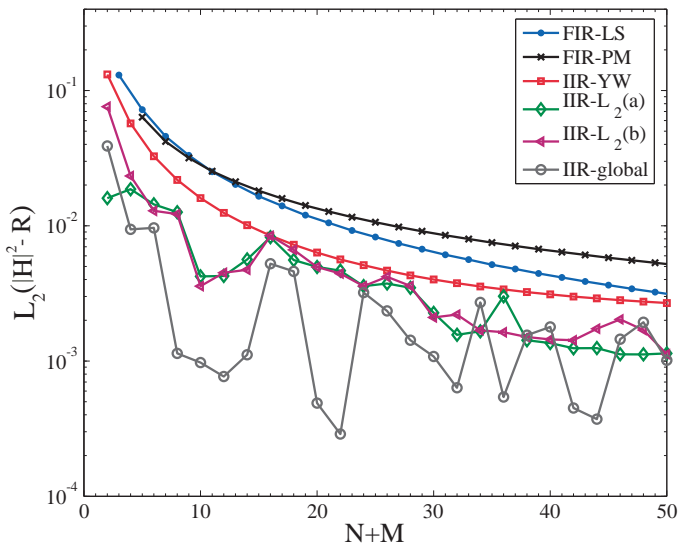}
\caption{Plot of $L_2$ error, $L_2(|H|^2 - R)$, vs. $N+M$
    for the Planck distribution $R(\omega)$ given in
    Fig.~1. The $L_2(|H|^2-R)$ error is plotted for six
    different filter methods: IIR Yule-Walker (red squares), FIR
    Parks-McClellan (black crosses), FIR Least-Square (blue dots), IIR
    global-optimum filter (gray circles), and two nonlinear
    conjugate-gradient methods optimizing two different $L_2$ norms:
    (a) $L_2(|H|-\sqrt{R})$ (green diamonds) and
    (b) $L_2(|H|^2-R)$ (magenta triangles).}
\end{center}
\label{fig:l2b}
\end{figure}

We suspect that the Chebyshev norm is typically the most appropriate
one for physical simulations involving random processes.  The reason
is that a large error in a narrow bandwidth, which might be allowed by
a least-square ($L_2$) norm, could result in a spectral feature that
might be mistaken for a spurious physical phenomenon; a large
``spike'' in error may also interact adversely with nonlinearities in
the physics.  Nevertheless, in Fig.~3 and Fig.~4, we show two
different $L_2$ errors, $L_2(|H|-\sqrt{R})$ and $L_2(|H|^2-R)$, for
the same filter designs as in Fig.~2, and the results demonstrate that
the Chebychev-optimal IIR filter is at least comparable, and often
superior to, the other methods, even in norms that it does not
strictly optimize. In particular, the $L_2(|H|^2-R)$ norm of the
Chebyshev-optimum IIR filter does better than all other local
optimization techniques for most $N+M$.  Further gains could
potentially be made, if this is the desired error norm, by starting
with the Chebyshev filter and then performing a local optimization by
some other method.

\section{Acknowledgements}

We are grateful to Alan V. Oppenheim, Stephen Boyd, and Almir Mutapcic
for helpful discussions. This work was supported in part by the
Materials Research Science and Engineering Center program of the
National Science Foundation under award DMR 02-13282, by a Department of
Energy (DOE) Computational Science Fellowship under grant
DE-FG02-97ER25308, and also by the Paul E. Gray Undergraduate Research
Opportunities Program Fund at MIT.


\end{document}